\documentclass[twocolumn,reprint, secnumarabic,amssymb, nobibnotes, aps, prd]{revtex4-1}

\usepackage{calc}
\usepackage{comment}
\usepackage{graphicx}
\usepackage{caption}
\usepackage{subcaption}
\usepackage[shortlabels]{enumitem}
\usepackage{amsmath}
\usepackage{amsfonts}
\usepackage{amssymb}
\usepackage{float}
\usepackage{xcolor}
\usepackage{spalign}
\usepackage{natbib}

\begin{document}

\title{Photoluminescence dynamics in few-layer InSe}

\author{Tommaso Venanzi$^{1,2}$}
\email[]{t.venanzi@hzdr.de} 
\author{Himani Arora$^{1,2}$}
\author{Stephan Winnerl$^{1}$}
\author{Alexej Pashkin$^{1}$}
\author{Phanish Chava$^{1,2}$}
\author{Amalia Patan{\`{e}}$^{3}$}
\author{Zakhar D. Kovalyuk$^{4}$}
\author{Zakhar R. Kudrynskyi$^{3}$}
\author{Kenji Watanabe$^{5}$}
\author{Takashi Taniguchi$^{5}$}
\author{Artur Erbe$^{1}$}
\author{Manfred Helm$^{1,2}$}
\author{Harald Schneider$^{1}$}

\affiliation{1 Helmholtz-Zentrum Dresden-Rossendorf, 01314 Dresden, Germany}
\affiliation{2 Technische Universit\"at Dresden, 01062 Dresden, Germany}
\affiliation{3 School of Physics and Astronomy, University of Nottingham, Nottingham NG7 2RD, UK}
\affiliation{4 Institute for Problems of Materials Science, The National Academy of Sciences of Ukraine, Chernivtsi, Ukraine}
\affiliation{5 National Institute for Material Science, 1-1 Namiki, Tsukuba, 305-0044, Japan}

\begin{abstract}

We study the optical properties of thin flakes of InSe encapsulated in hBN. More specifically, we investigate the photoluminescence (PL) emission and its dependence on sample thickness and temperature. Through the analysis of the PL lineshape, we discuss the relative weights of the exciton and electron-hole contributions. Thereafter we investigate the PL dynamics. Two contributions are distinguishable at low temperature: direct bandgap electron-hole and defect-assisted recombination. The two recombination processes have lifetime of $\tau_1 \sim 8\;$ns and $\tau_2 \sim 100\;$ns, respectively. The relative weights of the direct bandgap and defect-assisted contributions show a strong layer dependence due to the direct-to-indirect bandgap crossover. Electron-hole PL lifetime is limited by population transfer to lower-energy states and no dependence on the number of layers was observed. The lifetime of the defect-assisted recombination gets longer for thinner samples. Finally, we show that the PL lifetime decreases at high temperatures as a consequence of more efficient non-radiative recombinations.

\end{abstract}

\maketitle

\section{Introduction}

Van der Waals (vdW) semiconductors are a very interesting and promising class of materials. By stacking different vdW semiconductors on top of each other, it is possible to combine their properties \cite{Geim2013}. While a large number of material combinations is possible, the properties of each do not simply add up. There is a non-trivial interaction between different layers of the heterostructures. Already many interesting aspects of the physics of heterostructures have been discovered, for instance interlayer excitons \cite{Rivera2018,Merkl2019}, the interaction between excitons and the moir{\'{e}} potential \cite{Seyler2019,Tran2019}, and exciton condensation at high temperature \cite{Wang2019}. But even without considering the interest related to technological applications \cite{Liang2019}, there is still a lot of physics to investigate. In this vast set of vdW heterostructures, we focus here on hBN/InSe/hBN heterostructures and, more specifically, we study their optical properties analyzing the photoluminescence (PL) emission.

Few-layer InSe has shown promising properties for electronic applications because it features a direct bandgap and a high electron mobility (low electron effective mass) \cite{Mudd2016}. There are many studies demonstrating the possibility to use this material as an active layer for field-effect transistors \cite{Feng2014,Sucharitakul2015}, photodetectors \cite{Tamalampudi2014,Lei2014}, and optoelectronic devices for the infrared region \cite{Magorrian2018}. However, one issue that needs to be tackled is the contamination of thin-layer InSe when exposed to air \cite{Wei2018,Himani2017}. A solution for this issue is to embed the InSe flake in hBN. This encapuslation procedure protects the material from contamination and assures good optical and electronic properties, as shown recently for InSe itself and other vdW materials \cite{Bandurin2017,Lee2015,Himani2019}.

Here, we present an investigation of the photoluminescence emission from thin flakes of InSe of different thicknesses encapsulated in hBN. In particular, we investigate the temperature dependence of the PL emission analyzing the data using a modified version of Katahara's model \cite{Katahara2014}. Through this analysis we show that for thin-layer InSe the exciton interaction is observable only at low temperature. Moreover, we present layer-dependent time-resolved PL of InSe encapsulated in hBN and we determine the time constants of the radiative recombination. The PL decay shows two components that are the electron-hole and defect-assisted radiative recombination. The ratio of the weights of the two PL components varies with the number of layer, in agreement with the crossover from direct bandgap in the bulk to the indirect bandgap in few-layer crystals. Finally, a decrease of the PL lifetime is observed at higher temperature due to non-radiative scattering. 

\section{Sample and methods}

\begin{figure}
\centering
\includegraphics[scale=0.40]{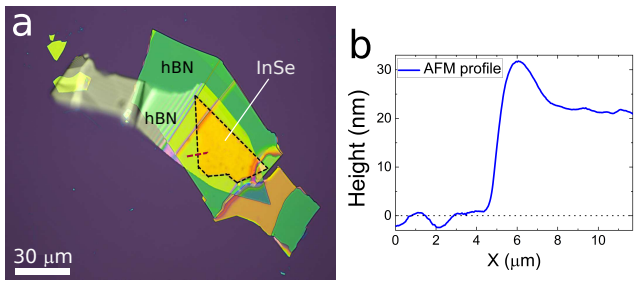}

\caption{(a) Optical image of an InSe flake fully-encapsulated in hBN. The black dashed line highlights the InSe flake. (b) AFM profile along the dashed red line in the optical image of the sample.}
\label{fig:sample}
\end{figure}

Figure \ref{fig:sample}a shows an optical image of a hBN-encapsulated InSe on SiO$_2$($285\;$nm)$/$Si substrate. The hBN-encapsulated InSe samples were fabricated using the dry transfer technique under ambient condition in an ISO 4 cleanroom environment, as described previously by L. Wang et al. \cite{Wang2013}. By using this polymer assisted fabrication technique, we ensured clean interfaces between InSe and hBN. This technique allows us to obtain samples with good crystal quality and low contamination. 

After the fabrication, the samples were kept in vacuum ($10^{-6}\;$mbar at $T = 300\;$K) in  order to prevent any degradation. For the time-integrated PL measurements we used a cw frequency-doubled Nd:YAG laser as excitation pump at a wavelength of $\lambda = 532\;$nm. The spot diameter on the sample was approximately $3$ $\mu$m, i.e. smaller than the heterostructure size. A nitrogen-cooled Si-CCD deep-depletion camera was used to detect the PL emission, which is dispersed in a spectrometer. For the time-resolved measurements, we used a single photon avalanche diode as detector coupled to a spectrometer. With this system $60\;$ps time resolution and $1\;$meV spectral resolution were obtained. A mode-locked Ti:Sa oscillator with pulse length of $3\;$ps in combination with a BBO frequency-doubler was used as excitation source at $\lambda=405\;$nm. The repetition rate was reduced from $78\;$MHz to $6.5\;$MHz by pulse-picking in a Pockels cell: this was done because of the relatively long PL lifetime of the sample. For time-resolved measurements we used a larger spot diameter (around $10\;\mu$m) in order to collect more PL emission, while keeping the excitation power density as low as possible.

The thicknesses of the encapsulated InSe flakes range from $20\;$nm to $2.4\;$nm, corresponding to around 24 atomic layers to 3 layers \cite{Mudd2016}. The thickness was measured using atomic force microscopy, as shown in figure \ref{fig:sample}b. We note that the hump in the AFM profile is due to a bubble at the edge between the InSe flake and top-encapsulating hBN layer. 

As shown recently \cite{Mudd2013,Mudd2016,Zheng2017,Guo2017,Hamer2019}, the band structure of InSe is strongly dependent on the number of layers. Bulk InSe has a direct bandgap while monolayer and few-layer InSe have an indirect bandgap. The difference in energy between direct and indirect bandgap is around $70\;$meV for monolayer InSe and it decreases rapidly upon increasing the number of layers. The direct bandgap is at the $\Gamma$ point. Upon decreasing the number of layers, the conduction band does not change qualitatively but the valence band forms a Mexican hat-like dispersion centered at the $\Gamma$ point so that thin layers of InSe have an indirect bandgap. The samples presented in this study cover the range of the direct-to-indirect bandgap crossover.

\section{Steady-state Photoluminescence}

\begin{figure*}
\centering

\includegraphics[scale=0.55]{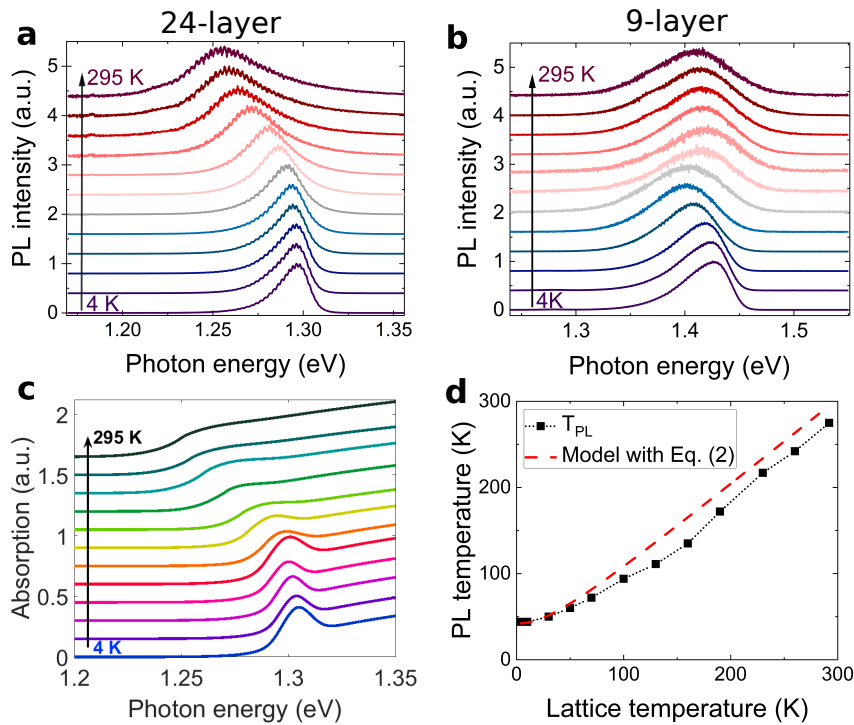}

\caption{(a) and (b) Temperature dependence of the PL spectra of a 24-layer and 9-layer InSe samples, respectively. The excitation power is $10\;\mu$W. For the 24-layer thick sample a redshift of the PL emission is observed due to the reduction of the bandgap energy. The 9-layer sample shows an s-shape due to the large contribution of the defect-assisted recombination. The temperatures of the PL spectra are the same as for the data points in figure (d). (c) Absorption spectra as extracted from the fitting of the PL spectra for the 24-layer thick sample. An exciton feature is observable at low temperature. (d) Effective PL temperature obtained from the model with respect to the actual lattice temperature.}
\label{fig:tempDepPL}
\end{figure*}

Figures \ref{fig:tempDepPL}a and \ref{fig:tempDepPL}b show temperature-dependent PL spectra for 24- and 9-layer InSe crystals, respectively. The energy position of the PL band of the two samples at $4\;$K is in agreement with previous reports on thin layers of InSe \cite{Mudd2014}. At $T=4\;$K both samples show a single broad PL emission band mostly due to defect-assisted radiative recombination. As noted in \cite{Mudd2013}, the large broadening of the PL lines is due to the low electron mass that makes the emission very sensitive to any surface effect and to any disordered potential. The broadening of the PL lines increases while decreasing the number of layers, as becomes clear from the broad emission of the 9-layer sample.

While increasing the temperature, the 24-layer sample shows a monotonic redshift that is a consequence of the reduction of the bandgap energy due to the lattice expansion and to the interaction with phonons. Further details are given in supplementary material (SM) \cite{Venanzi2019s}. The temperature dependence of the PL energy position of the 9-layer flake shows an s-shape. This is a consequence of the defect-state emission that dominates the PL at low temperature, as discussed in previous studies and observed in other semiconductor systems \cite{Mudd2014,Cho1998}. The PL energy position of the 9-layer sample shows an overall blueshift in comparison with the thicker InSe sample. This is due to quantum confinement and is extensively reported in the literature \cite{Mudd2013,Zheng2017}. 

To get more information on the PL emission mechanisms, we looked at the PL lineshape of the 24-layer sample. We consider the model proposed by Katahara et al. \cite{Katahara2014} for PL emission and we adapt it to our case. This model is a generalized version of the van Roosbroeck-Shockley equation that connects absorption and interband PL in semiconductors \cite{VanRoosbroeck1954,Lasher1964}. This means that we can extract absorption from PL data. Moreover we consider both band-to-band and exciton absorption. In this way we can evaluate the relative weight of these two contributions. The expression used for modeling the PL lineshape is:
\begin{equation}
I_{PL}(E) \propto \frac{E^2 a(E)}{\exp(\frac{E-\Delta\mu}{kT_{PL}})-1}\cdot\bigg(1-\frac{2}{\exp(\frac{E-\Delta\mu}{2kT_{PL}})+1}\bigg)
\label{eq:fit}
\end{equation}
where the first part is the connection between Planck's law and the absorption, and the second part in bracket is a small correction that takes into account the occupation of the bands (Pauli blocking). Here, $a(E) = a_B(E) + p \;a_X(E)$ is the total absorption given by a linear combination of band-to-band and exciton contribution, $\Delta\mu$ is the quasi-Fermi energy, and $T_{PL}$ is the effective photoluminescence temperature. We note that in many cases the part of the expression containing the temperature can be simplified assuming a Boltzmann distribution. The quasi-Fermi energy is an effective Fermi energy introduced by Lasher et al. in order to consider the occupation of the conduction and valence bands and it assumes values close to the bandgap energy \cite{Lasher1964}. The model is reported in detail in SM \cite{Venanzi2019s}.

Figure \ref{fig:tempDepPL}c shows the absorption spectra deduced from equation \ref{eq:fit}. At low temperatures the exciton resonance is clearly observable in the absorption spectra and it smears out at higher temperature as expected. In fact, the exciton binding energy is around $14\;$meV \cite{Merle1978,Camassel1978,Shubina2019} that corresponds to about $160\;$K.

We want to highlight another detail of our fitting procedure. We introduced an effective PL temperature to reproduce accurately the PL spectra especially at low temperature. This is strictly connected to the inhomogeneous broadening of the PL line. In fact the effective PL emission temperature is determined by the high-energy side of the PL emission, where the excited carriers can thermalize. More specifically the effective PL temperature is proportional to the slope of the exponential high-energy tail of the PL emission \cite{Yoon1996,Schnabel1992}. The high energy tail is due to the thermal population of the bands that is in very good approximation due to a Boltzmann distribution. The higher the PL temperature, the less steep is the exponential decay of the PL high energy tail. This effect can be observed in figure \ref{fig:tempDepPL}a. At $4\;$K one would expect a very sharp decay, but we observe a smoother decay due to inhomogeneous broadening. Figure \ref{fig:tempDepPL}d shows the PL temperature versus the lattice temperature. The PL temperature does not approach zero Kelvin but it saturates around a certain value much larger than zero.

We can model this behavior with this formula:
\begin{equation}
T_{PL} = \sqrt{T^2+T_0^2},
\end{equation}
in analogy with the model proposed by Marianer et al. for disordered semiconductors \cite{Marianer1992,Venanzi2019}. The idea is to incorporate the disorder, that in our case shows up as inhomogeneous broadening of the PL line, in the effective temperature. This simple model reproduces the observed electron-hole temperature reasonably well. Therefore, $T_0 = 44\;$K$=4\;$meV is a measurement of the disorder in the material and it can be intuitively interpreted as the standard deviation of the disorder potential in the sample due to lattice defects and sample inhomogeneity. The PL temperature at zero Kelvin $T_0$ increases upon decreasing the number of layer (see SM \cite{Venanzi2019s}). This behavior indicates that the disorder is more pronounced for thinner samples, as expected. This treatment has a general validity for semiconductors with direct bandgap and parabolic dispersion.
 
We finally note that a non-vanishing $T_0$ could be due to higher electron-hole temperature induced by laser excitation. However, in our case this effect is negligible because we do not observe a dependence of the parameter $T_0$ on excitation power (see SM \cite{Venanzi2019s}). 

\section{Time-resolved photoluminescence}

\begin{figure*}[]

\centering
\includegraphics[scale=0.81]{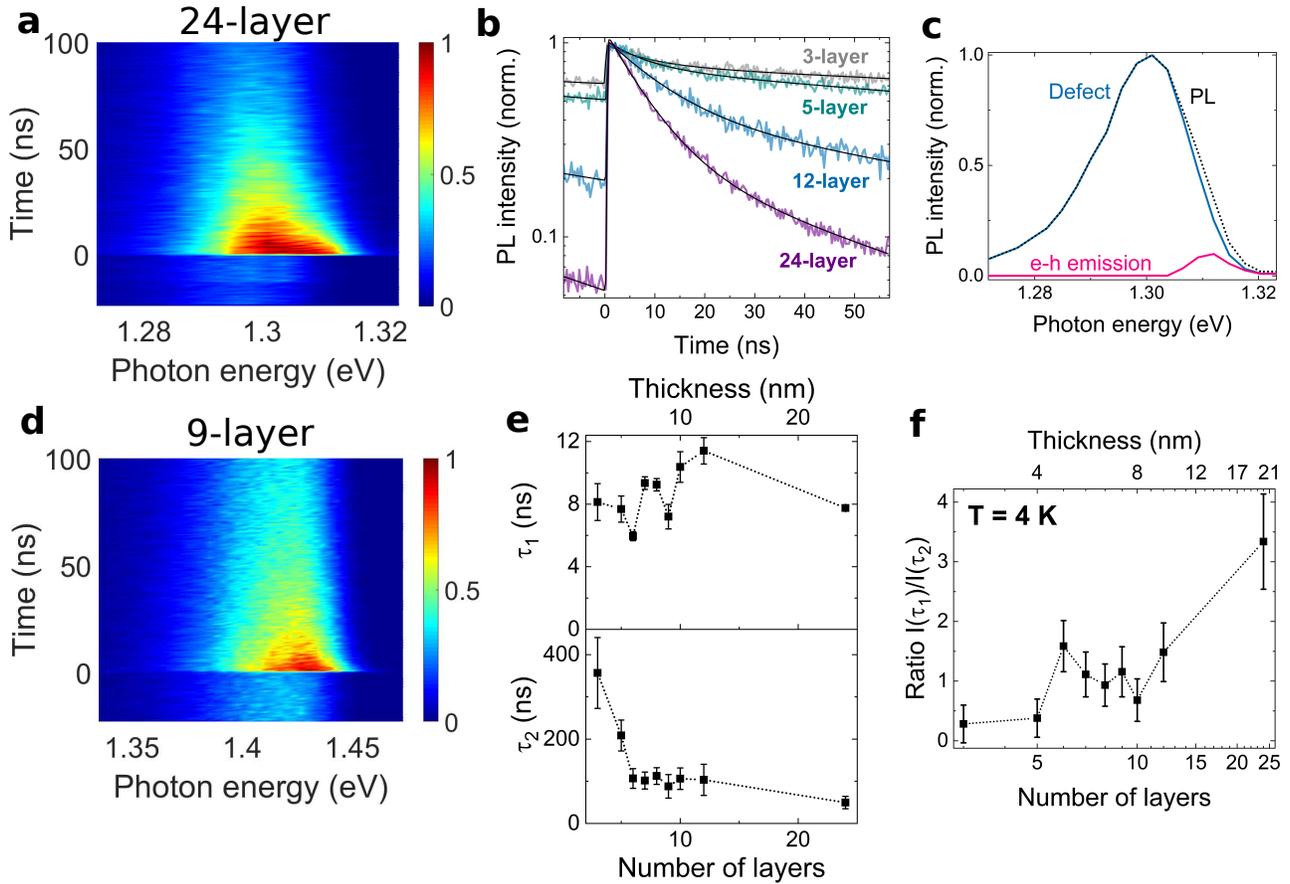}

\caption{(a) and (d) 2D false-color maps of normalized PL intensity as a function of time and photon energy at $T=4\;$K and $10\;\mu$J$\;$cm$^{-2}$ excitation fluence. The two maps are of a 24- and 9-layer sample, respectively. (b) PL decays integrated in photon energy of samples with different thickness. The decays are bi-exponential. (c) Spectra of the extracted fast (electron-hole) and slow (defect) components for the 24-layer thick sample. (e) Lifetimes of the fast and slow components of the PL decay as a function of sample thickness. (f) Dependence of the ratio between PL contributions associated with the fast and the slow component on the number of layers.}
\label{fig:trplMap}
\end{figure*}

Figure \ref{fig:trplMap}a shows the PL intensity as a function of time and photon energy for a 24-layer InSe crystal. The PL emission shows a spectral dependent bi-exponential decay. In order to get the time constants, we average the PL emission over the photon energy and we fit the data with a bi-exponential decay convoluted with the instrument response function. The energy-integrated PL decay is shown in figure \ref{fig:trplMap}b. The fast and slow PL components have lifetime of $\tau_1=7.7\pm0.2\;$ns and $\tau_1=49\pm6\;$ns, respectively. In order to visualize the two PL components independently, 2D false-color plots of the extracted fast and slow decays are shown in supplementary material \cite{Venanzi2019s}. 

The slow PL decay is a fingerprint of a defect-assisted electron-hole recombination. Therefore we associate the slow component with this radiative channel. The fast component is associated with the direct-bandgap electron-hole recombination. 

Integrating in time the fast and the slow components as obtained from the fit, it is possible to separate the spectra of the two contributions (shown in figure \ref{fig:trplMap}c). The central emission energy of the fast component is $1.312\pm 0.001\;$eV with full width at half maximum (FWHM) of $6.8\pm 0.5\;$meV, while the slow component is centered at $1.300\pm 0.001\;$eV with FWHM of $15\pm 1\;$meV. The slow component lies at lower energy and with a broader spectrum, as expected for a defect-assisted emission. The two components were not spectrally distinguishable in steady-state PL.

Now we look at the layer dependence of the time-resolved PL. Figure \ref{fig:trplMap}d shows the PL data for a 9-layer thick InSe crystal measured under the same condition as the 24-layer thick sample. Qualitatively, it is evident that the PL decay takes place on a longer timescale. In addition, the weights of the two PL components change significantly. The defect-assisted recombination appears to be more dominant with respect to the electron-hole recombination for thinner samples, as expected from the direct-to-indirect bandgap crossover driven by the sample thickness \cite{Mudd2013}. Conversely, electrons and holes in the 24-layer sample can recombine easily because of the direct bandgap, while in thin samples they need the assistance of a scattering process.

\begin{figure*}[]

\centering
\includegraphics[scale=0.61]{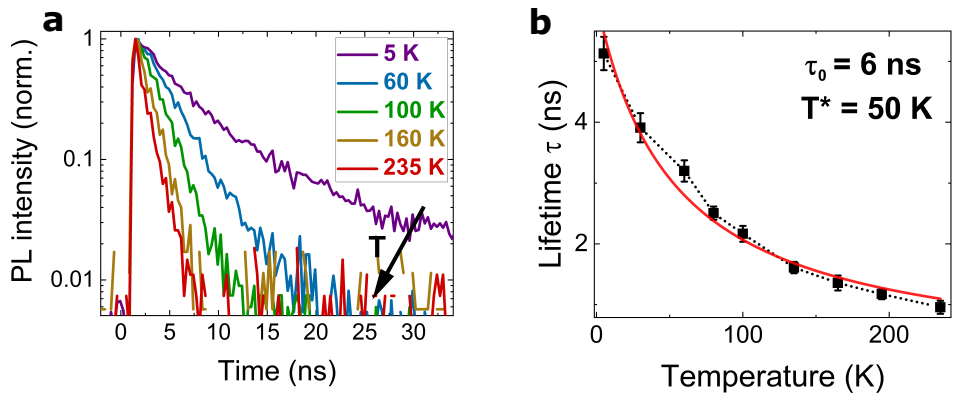}

\caption{(a) PL decays as a function of temperature. The PL decay gets faster at higher temperature due to the more efficient non-radiative scattering. The sample is 24-layer thick and fully encapsulated in hBN. (b) PL lifetime $\tau_1$ as a function of temperature. The red curve is obtained from equation \ref{eq:tau}.}
\label{fig:trplVsT}
\end{figure*}

To further corroborate this observation and to perform a quantitative analysis, we fabricated and measured a series of encapsulated InSe crystals with different thicknesses. Figure \ref{fig:trplMap}b shows the PL decays integrated in photon energy for different layer thicknesses. 

We fitted the PL decay with a bi-exponential decays and we extracted the fast and slow lifetimes ($\tau_1$ and $\tau_2$) for each sample. The lifetimes are shown in figure \ref{fig:trplMap}e. We do not observe a clear layer-dependence of the fast component, i.e. electron-hole recombination. On the other hand, the lifetime of the slow component increases with decreasing the flake thickness. The dynamics of the electron-hole recombination is limited at low temperature mainly by the population transfer to lower energy states, i.e. bound states below the bandgap. This could explain why we do not observe a layer dependence for the fast component of the decay. 

The ratio $\frac{I(\tau_1)}{I(\tau_2)}$ for each sample is shown in figure \ref{fig:trplMap}f. The fast component becomes more and more dominant with increasing the number of layers. The increase of the ratio takes place between $5$ and $10$ layers, i.e. in good agreement with the number of layers where the direct-to-indirect bandgap crossover is expected \cite{Mudd2016}. The error bars were obtained from the standard errors of the least-square fitting procedure and the propagation of errors.

Finally, we consider the temperature dependence of the PL dynamics. Figure \ref{fig:trplVsT}a shows the electron-hole PL decays as a function of temperature. The PL lifetime decreases monotonically with increasing the temperature due to higher non-radiative scattering, i.e. phonon scattering.  We note that the PL decays at temperature higher than $70\;$K show a single exponential decay. This is because the defect states responsible for the slow component are not stable anymore and the only radiative channel is the fast electron-hole recombination. 

Figure \ref{fig:trplVsT}b shows the extracted PL lifetime as a function of temperature. We use an empirical model in order to fit the data considering the radiative and non-radiative decay. Assuming an exponential decay, the lifetime is inversely proportional to the sum of the radiative and non-radiative decay rates:
\begin{equation}
\tau = \frac{1}{\beta_r + \beta_{nr}} = \frac{1}{\beta_0 + \alpha_{nr} T}
\label{eq:tau}
\end{equation}
where $\beta_r$ and $\beta_{nr}$ are, respectively, the radiative and non-radiative decay rates, $\beta_0$ is the decay coefficient at zero Kelvin, and a linear approximation was done to model the dependence of the non-radiative scattering on temperature. 

This simple approximation shows a good agreement with the experimental data and gives two quantitative information: 1) the lifetime at zero Kelvin ($\frac{1}{\beta_0}=\tau_0=6\;$ns) and 2) the temperature $T^*$ at which the non-radiative scattering overcomes the radiative scattering. The latter is: $T^* = \frac{\beta_0}{\alpha_{nr}}=50\;$K. 

Finally, we note that if we use a square root temperature dependence in equation \ref{eq:tau} instead of a linear one, the model looks like the Schockley-Read-Hall model for non-radiative scattering \cite{Schockley1952,Connelly2010}. However, this gives a worse agreement with the experimental data and no significant parameters can be extracted. We also note that, according to the data, the best-fitting exponent for the temperature would be $1.34\pm0.06 \thickapprox \frac{4}{3}$ instead of $1$, but no physical explanation could be found.

\section{Conclusion}

We investigated the PL and time-resolved PL emission from few-layer InSe fully-encapsulated in hBN. A lineshape analysis of the time-integrated emission reveals the contribution of exciton and electron-hole recombination. The excitonic emission was observed only at low temperatures. The role of disorder in the material can be modeled very well by introducing an effective temperature.

The analysis of the time-resolved PL signals allows us to disentangle the contribution from direct bandgap electron-hole recombination ($ \tau_1 \sim 8\;$ns) and to defect-assisted recombination ($\tau_2 \sim 100\;$ns). These contributions are not spectrally distinguishable without resolving the dynamics. The defect-assisted contribution becomes increasingly important as the number of layers is decreased. Remarkably, the electron-hole PL lifetime basically is independent of the number of layers, while the lifetime of the defect-assisted recombination increases for thinner samples. Furthermore, shorter PL lifetimes were found with increasing temperature, which is caused by more efficient non-radiative recombination. 

In summary, the analysis of a comprehensive set experimental data on the PL emission dynamics of few-layer InSe encapsulated in hBN allows us to distinguish the involved microscopic physical mechanisms. These results are important for technological applications based on few-layer InSe.

\section*{Acknowledgement}

The authors cordially thank Pedro Pereira for giving friendly help during the course of this work. This work has been partially funded by the Initiative and Networking Fund of the German Helmholtz Association, Helmholtz International Research School for Nanoelectronic Networks NanoNet (VH-KO-606). We acknowledge the European Union’s Horizon 2020 research and innovation programme Graphene Flagship Core 2 under grant agreement number 785219. We acknowledge also the National Academy of Sciences of Ukraine. Growth of hexagonal boron nitride crystals was supported by the Elemental Strategy Initiative conducted by the MEXT, Japan and the CREST(JPMJCR15F3), JST.

\end{document}